\documentclass[a4paper,11pt]{article}
\usepackage{pos}

\title{Role of meson exchange and final-state interaction in $J/\psi$-meson
photoproduction}

\author*[a]{Sang-Ho Kim}
\affiliation[a]{Department of Physics and
Origin of Matter and Evolution of Galaxies (OMEG) Institute,\\
Soongsil University, Seoul 06978, Republic of Korea}

\emailAdd{shkimphy@gmail.com}
\abstract{
We investigate $J/\psi$ photoproduction off the nucleon using a dynamical model based
on a Hamiltonian framework.
First, Pomeron exchange is regarded as a background contribution that accounts for the
gradual rise of the total cross section with increasing beam energy.
Then, the meson-exchange and direct $J/\psi$-radiation contributions are included as
part of the Born term.
More specifically, we examine the contributions of the light-meson [$\pi^0(135)$,
$\eta(548)$, $\eta'(958)$, $f_1(1285)$] and charmonium-meson [$\eta_c(1S)$,
$\chi_{c0}(1P)$, $\chi_{c1}(1P)$, $\eta_c(2S)$, $\chi_{c1}(3872)$] exchanges in the
$t$-channel diagram.
Most of the coupling constants are determined from the radiative decays of
the $J/\psi$ or other relevant charmonium states.
Finally, we consider the $J/\psi N$ final-state interaction (FSI) within the
leading-order approximation, which includes gluon-exchange and direct
$J/\psi N$-coupling terms.
The model parameters of the Hamiltonian are fitted to the data from the GlueX and
$J/\psi$-007 experiments at Jefferson Laboratory (JLab).
The light mesons $\eta(548)$ and $\eta'(958)$ are found to make the most significant
contributions, whereas the charmonium mesons play only a minor role.
The direct $J/\psi$-radiation term begins to come into play only in the backward
scattering regions.
The FSI contribution is suppressed by a factor of $10^1 - 10^2$ in comparison to the
Born-term contribution when a Yukawa-type potential is employed for the
charmonium-nucleon interaction.
Our results agree well with the available JLab data on the total and $t$-dependent
differential cross sections.
High-precision, angle-dependent data in the threshold region ($E_\gamma \leqslant
8.9$ GeV) are crucial for confirming the FSI effect.
}

\FullConference{The 21st International Conference
on Hadron Spectroscopy and Structure (HADRON2025)\\
27 - 31 March, 2025\\
Osaka University, Japan\\}

\begin{document}
%-------------------------------------------------------------------------------------
\maketitle

%=====================================================================================
\section{Introduction and formalism}
%=====================================================================================

In recent years, considerable progress has been made in the study of $J/\psi$
photoproduction off the nucleon, particularly since the first near-threshold
exclusive data became available from Jefferson Laboratory
(JLab)~\cite{GlueX:2019mkq}.
Pronounced cusp structures are observed at the $\bar D \Lambda_c$ and $\bar D^*
\Lambda_c$ thresholds~\cite{GlueX:2019mkq,GlueX:2023pev,Du:2020bqj}, and the
existence of $s$-channel hidden-charm pentaquark states $P_c$ remains under
debate~\cite{LHCb:2015yax,LHCb:2019kea}.
The gravitational density of gluons is also investigated using various models through
threshold measurements of $J/\psi$ photoproduction~\cite{Duran:2022xag}.
In this work, we summarize our recent work on $J/\psi$ photoproduction using a
dynamical model based on a Hamiltonian framework~\cite{Kim:2025oyo}.
Our previous model for $\phi$ photoproduction~\cite{Kim:2021adl,Kim:2024lis} is
extended to the case of $J/\psi$ photoproduction.

\begin{figure*}[ht]
\centering
\includegraphics[width=14.0cm]{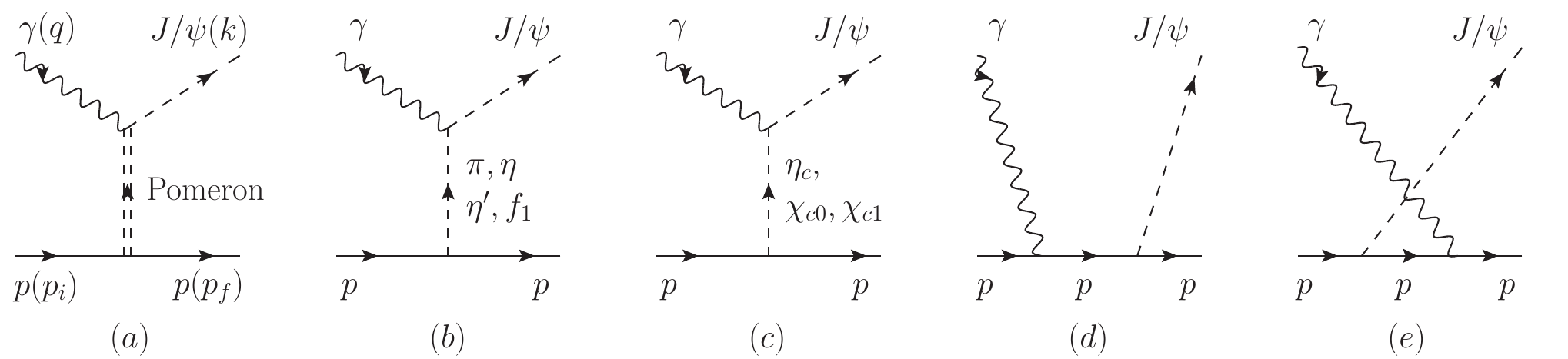}
\caption{Feynman diagrams for $\gamma p \to J/\psi p$,
which include the exchange of ($a$) Pomeron ($b$) light mesons, ($c$) charmonium
mesons in the $t$ channel, and $(d,\,e)$ direct $J/\psi$-radiation terms in the $s$-
and $u$-channels.}
\label{FIG01}
\end{figure*}

The Feynman diagrams are depicted in Fig.~\ref{FIG01}.
We rigorously examine the relative contributions of the light-meson [$\pi^0(135)$,
$\eta(548)$, $\eta'(958)$, $f_1(1285)$] and charmonium-meson [$\eta_c(1S)$,
$\chi_{c0}(1P)$, $\chi_{c1}(1P)$, $\eta_c(2S)$, $\chi_{c1}(3872)$] exchanges, as shown
in Fig.~\ref{FIG01}(b) and Fig.~\ref{FIG01}(c), respectively.

The interaction vertices are derived from the effective Lagrangians.
For example, the electromagnetic (EM) interactions relevant to the $t$-channel
diagram read
%EQUATION>>>
\begin{align}
\mathcal{L}_{\gamma \Phi J/\psi} = \frac{e g_{\gamma \Phi J/\psi}}{M_{J/\psi}}
\epsilon^{\mu\nu\alpha\beta} \partial_\mu A_\nu \partial_\alpha \psi_\beta \Phi,
\,\,\,
\mathcal{L}_{\gamma S J/\psi} = \frac{eg_{\gamma S J/\psi}}{M_{J/\psi}}
F^{\mu\nu} \psi_{\mu\nu} S,
\end{align}
%EQUATION<<<
and the strong interaction Lagrangians are expressed as
%EQUATION>>>
\begin{align}
\mathcal{L}_{\Phi NN} = -ig_{\Phi NN} \bar N \Phi \gamma_5 N ,
\,\,\,
\mathcal{L}_{S NN} = -g_{S NN} \bar N S N,
\end{align}
%EQUATION<<<
for the pseudoscalar ($\Phi$)- and scalar ($S$)-meson exchanges, respectively.
The relevant coupling constants appearing in the eﬀective Lagrangians are determined
from the radiative decays of $J/\psi$ and other charmonium mesons~\cite{PDG:2024cfk}.
Meanwhile, the meson-$NN$ coupling constants are taken from the Nijmegen
potentials~\cite{Stoks:1999bz,Stoks:1999bz2} and from the radiative decays of
charmonium mesons to $p \bar p$~\cite{PDG:2024cfk}, for the light- and
charmonium-meson exchanges, respectively.
The effective Lagrangians relevant to the $s$- and $u$-channel diagrams are defined
by
%EQUATION>>>
\begin{align}
\mathcal{L}_{\gamma NN} = - e \bar N
\left[ \gamma_\mu - \frac{\kappa_N}{2M_N} \sigma_{\mu\nu} \partial^\nu
\right] N A^\mu ,
\,\,\,
\mathcal{L}_{J/\psi NN} = - g_{J/\psi NN} \bar N
\left[ \gamma_\mu - \frac{\kappa_{J/\psi NN}}{2M_N} \sigma_{\mu\nu} \partial^\nu
\right] N \psi^\mu.
\end{align}
%EQUATION<<<

Note that the scattering amplitudes derived from the effective Lagrangians are, by
construction, gauge invariant for each contribution in the $t$-channel process.
However, the electric terms of the $s$- and $u$-channel amplitudes do not
individually satisfy the gauge invariance, but their sum does.

%FIGURE>>>
\begin{figure}[t]
\centering
\includegraphics[width=9.0cm]{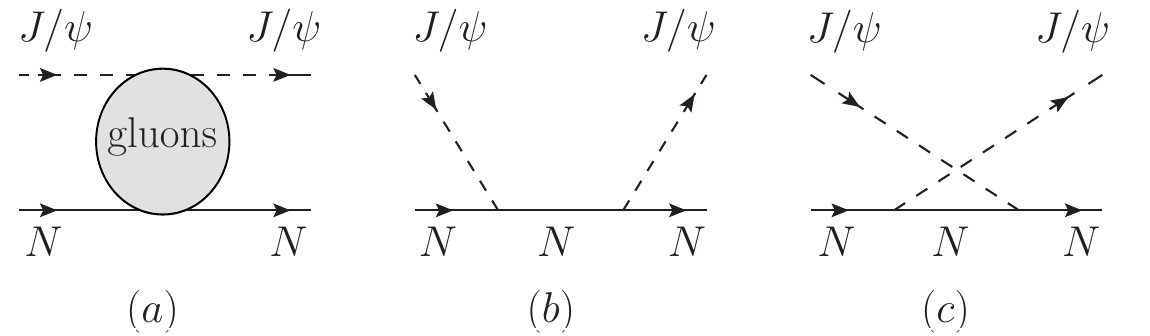}
\caption{$J/\psi N \to J/\psi N$ scattering amplitudes: (a) gluon-exchange, (b,c)
direct $J/\psi$-coupling terms.}
\label{FIG02}
\end{figure}
%FIGURE<<<
As for the $J/\psi N$ FSI amplitude, the gluon-exchange [Fig.~\ref{FIG02}(a)] and
direct $J/\psi$-coupling [Fig.~\ref{FIG02}(b,c)] terms are extracted by retaining only
the leading term in the Lippmann-Schwinger equation.
Given that lattice QCD calculation for the $J/\psi$ potential is available, we adopt
the form proposed in Ref.~\cite{Kawanai:2010ev} for the charmonium–nucleon potential,
which is found to be approximately of the Yukawa type
$( -v_0 \frac{e^{-\alpha r}}{r} )$.
The parameters are determined to be
%EQUATION>>>
\begin{align}
v_0 = 0.10,\,\,\, \alpha = 0.6\, {\rm GeV}~\cite{Kawanai:2010ev},\,\,\,
v_0 = 0.42,\,\,\, \alpha = 0.6\, {\rm GeV}~\cite{Brodsky:1989jd}.
\label{eq:yukawa}
\end{align}
%EQUATION<<<
%We attempt to use both in our calculation and compare the results.

%=====================================================================================
\section{Numerical Results}
%=====================================================================================

%FIGURE>>>
\begin{figure}[ht] % [h!] [p]
\centering
\includegraphics[width=6.5cm]{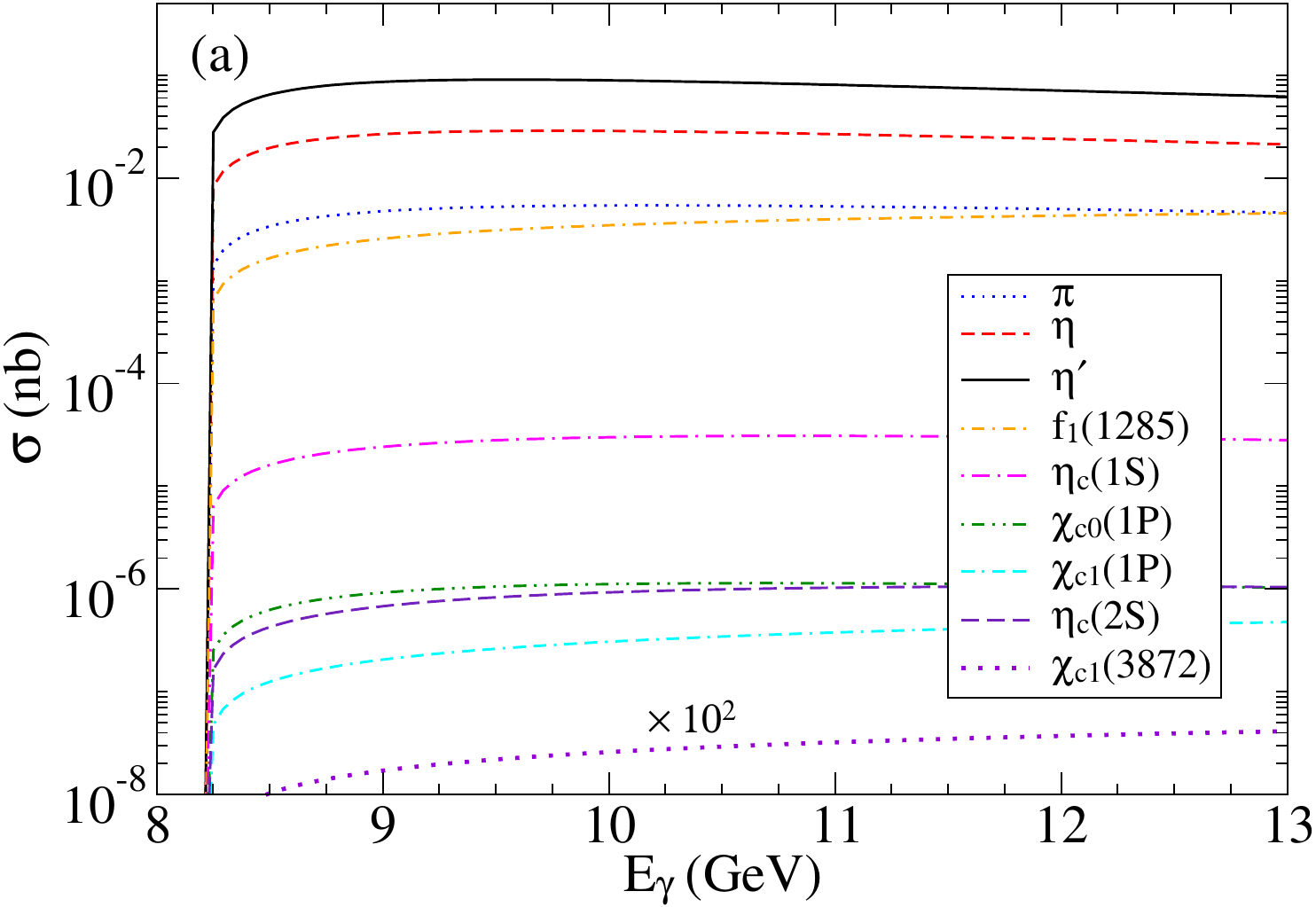} \hspace{1em}
\includegraphics[width=6.5cm]{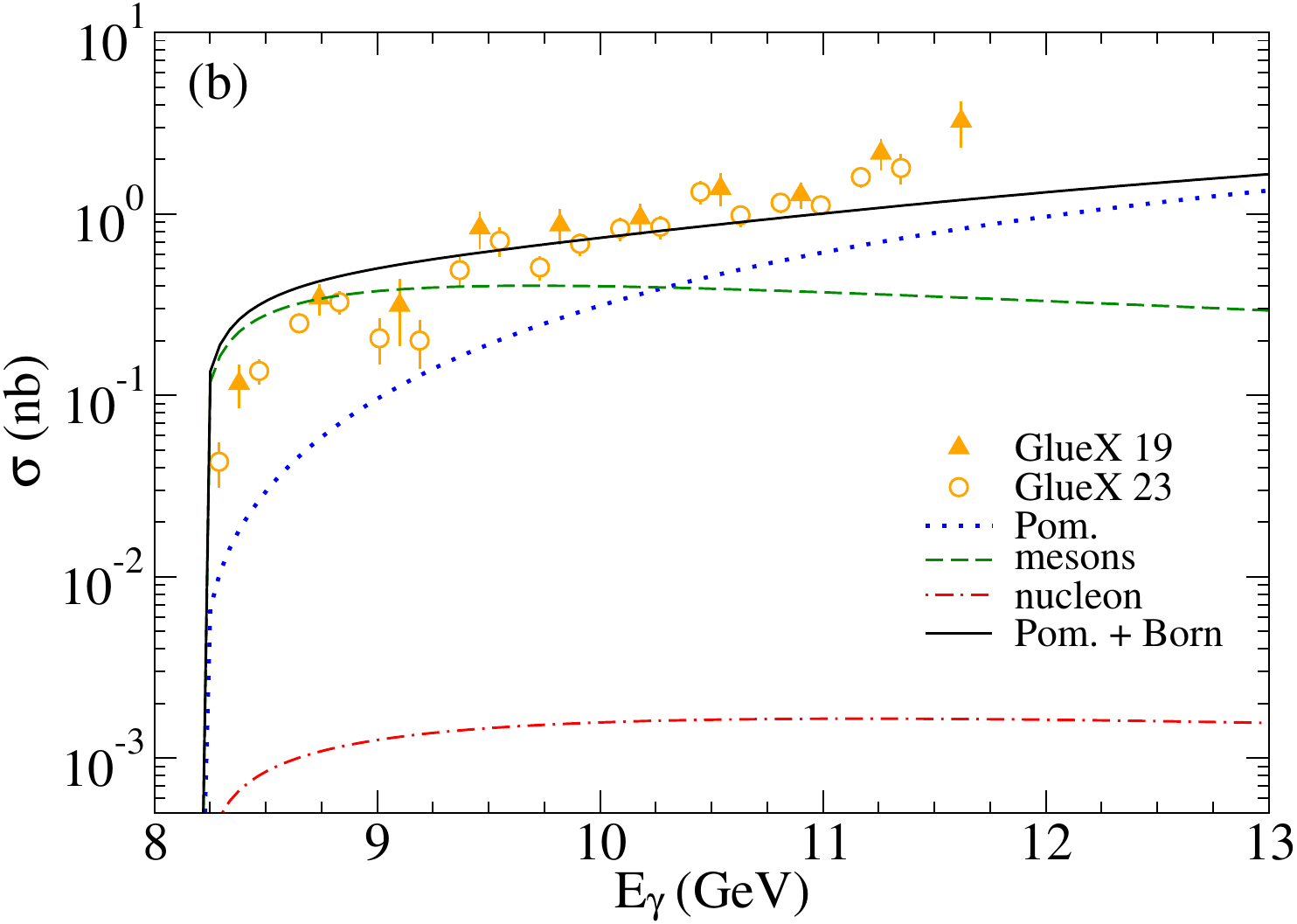}
\caption{(a) Each contribution of meson exchanges for for $\gamma p \to J/\psi p$.
(b) Total cross section without the FSI effect.
The GlueX~\cite{GlueX:2019mkq,GlueX:2023pev} data are used.}
\label{FIG03}
\end{figure}
%FIGURE<<<
In Fig.~\ref{FIG03}(a), we present each contribution of meson exchanges for
$\gamma p \to J/\psi p$.
We find that the differences in the cross sections between the light-meson and
charmonium–meson contributions are sufficiently large to be distinguished.
Indeed, the light mesons provide a more significant contribution than the charmonium
mesons.
The propagator, [1/($t$ - $M^2$)], in the scattering amplitude is primarily
responsible for this difference.
The dominant contributions from the light- and charmonium-meson exchanges are found
to arise from the $\eta' (958)$  and $\eta_c (1S)$ mesons, respectively.
Nevertheless, the $\eta_c (1S)$-meson exchange is suppressed by approximately three
orders of magnitude compared to the $\eta' (958)$-meson exchange. 

Figure~\ref{FIG03}(b) shows the total cross section without the FSI effect (black
solid), which includes the Pomeron exchange and the sum of meson exchanges.
The direct $J/\psi$-radiation contribution is also presented.
The results imply that the meson-exchange contribution is indeed essential, along
with the Pomeron exchange, for describing the low-energy GlueX
data~\cite{GlueX:2019mkq,GlueX:2023pev}.
Meanwhile, the direct $J/\psi$ radiation is suppressed by a factor of $10^2$ compared
to the meson-exchange contribution.

%FIGURE>>>
\begin{figure*}[t]
\centering
\includegraphics[width=13.5cm]{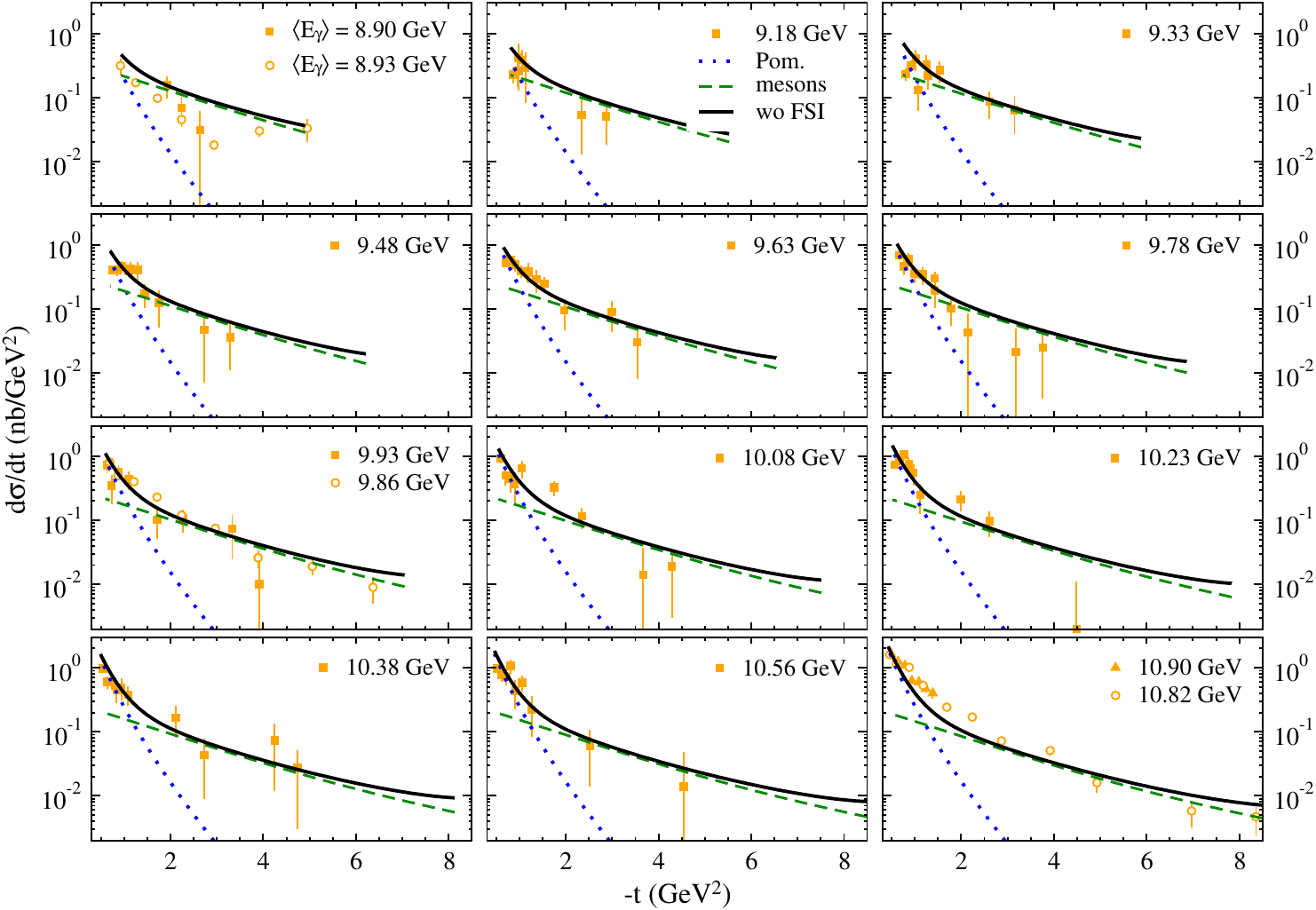}
\caption{Differential cross sections without the FSI effect.
The GlueX19~\cite{GlueX:2019mkq} (triangle), GlueX23~\cite{GlueX:2023pev} (circle),
and $J/\psi$-007~\cite{Duran:2022xag} (quadrangle) data are used.}
\label{FIG04}
\end{figure*}
%FIGURE<<<
The results for the differential cross sections are shown in Fig.~\ref{FIG04} as
functions of $-t$.
The Pomeron exchange alone yields a distribution that is too steep to account for the
available JLab data~\cite{GlueX:2019mkq,GlueX:2023pev,Duran:2022xag}.
The inclusion of meson exchanges significantly improves the agreement across the
entire range of scattering angles, as the slope of the meson-exchange contribution
closely matches that of the JLab data for $-t \geqslant 2~{\rm GeV}^2$.
The direct $J/\psi$-radiation term begins to play a role in the backward scattering
region, as expected.
Specifically, its contribution increases the Born term to some extent in the region
$-t \geqslant 5$ ${\rm GeV}^2$ compared to the meson-exchange contribution.
The GlueX23 data~\cite{GlueX:2023pev} cover the full range of scattering angles and
therefore provide a constraint on the direct $J/\psi$-radiation contribution.

%FIGURE>>>
\begin{figure}[h] %[h!]
\centering
\includegraphics[width=6.5cm]{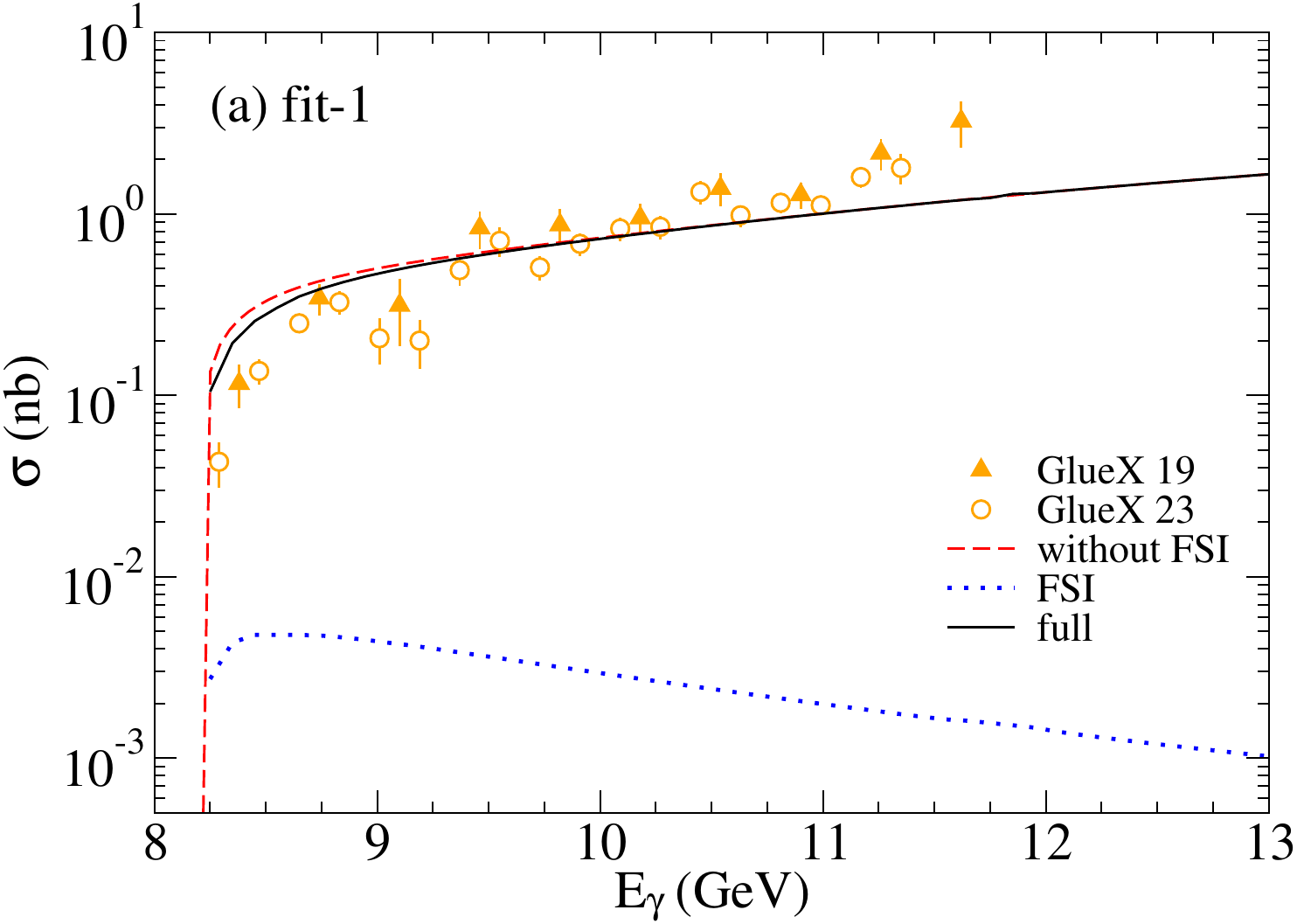} \hspace{1em}
\includegraphics[width=6.5cm]{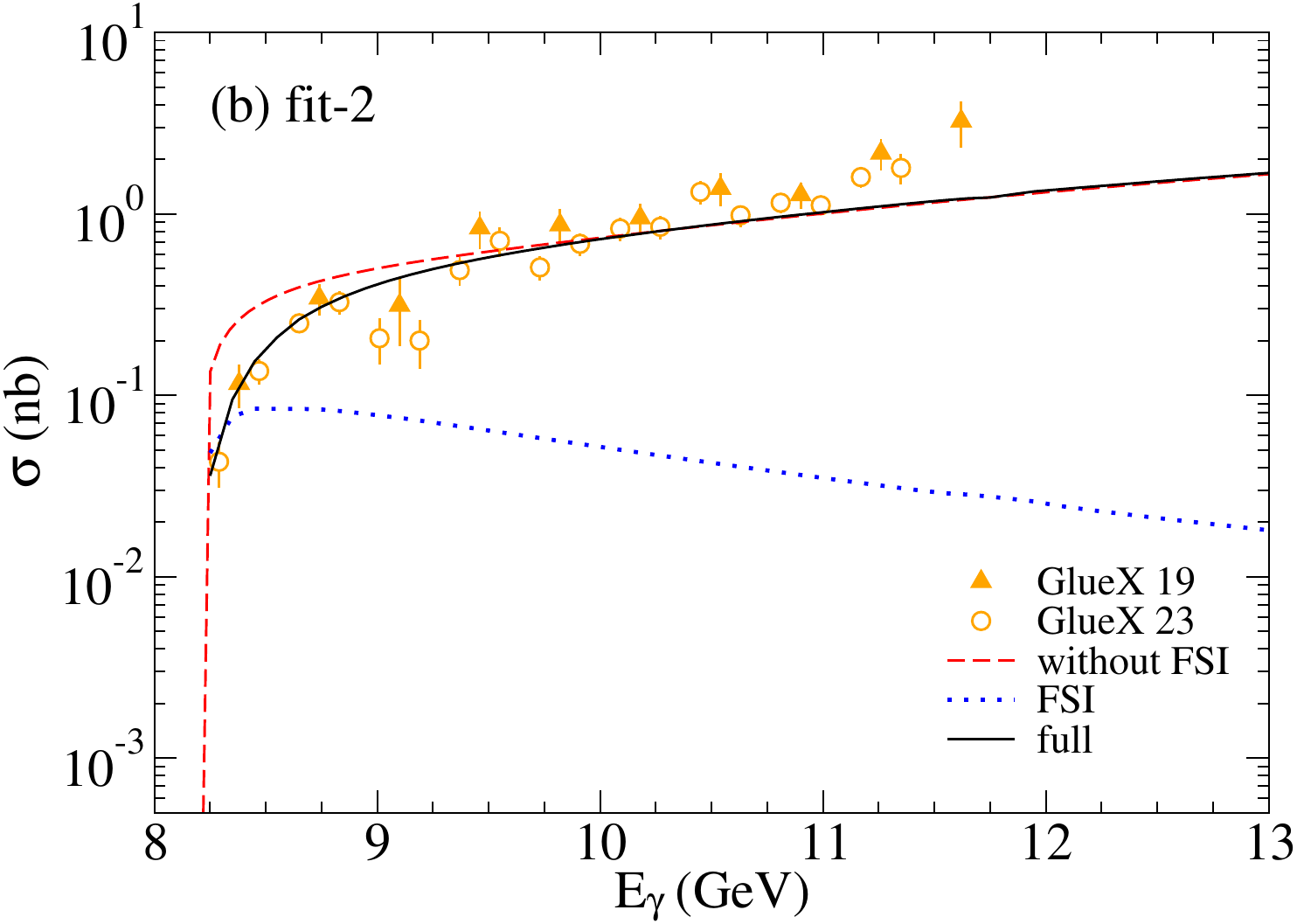}
\caption{Total cross section for $\gamma p \to J/\psi p$ including the FSI effect.
The Yukawa potential ($v_0$, $\alpha$) parameters used are:
(a) fit-1: (0.10, 0.6)~\cite{Kawanai:2010ev},
(b) fit-2: (0.42, 0.6)~\cite{Brodsky:1989jd}.
The GlueX data~\cite{GlueX:2019mkq,GlueX:2023pev} are used.}
\label{FIG05}
\end{figure}
%FIGURE<<<
We now present the result for the total cross section with the FSI effect included.
The contribution from the FSI term is found to originate entirely from the
gluon-exchange interaction [Fig.~\ref{FIG02}(a)], while the direct $J/\psi$-coupling
term [Fig.~\ref{FIG02}(b,c)] is relatively much more suppressed.
For the Yukawa potential in the gluon-exchange interaction, we employ two different
sets of parameters [see Eq.~(\ref{eq:yukawa})], which we refer to as the
fit-1~\cite{Kawanai:2010ev} and fit-2~\cite{Brodsky:1989jd} models, respectively.
In Fig.~\ref{FIG05}, we compare the results with and without the FSI effect.
The red dashed curve (sum of the Pomeron-exchange and Born terms) corresponds to the
black solid curve in Fig.~\ref{FIG03}(b), and it interferes destructively with the FSI
contribution.
The FSI effect is minor when the fit-1 model is used [Fig.\ref{FIG05}(a)], but
becomes significant with the fit-2 model [Fig.~\ref{FIG05}(b)].
Indeed, we observe a noticeable improvement near the threshold when using the fit-2
model.

%=====================================================================================
\section{Summary and Conclusions}
%=====================================================================================

We have investigated $J/\psi$-meson photoproduction off the nucleon target within a
dynamical model based on a Hamiltonian that generates both the $\gamma N \to J/\psi
N$ production amplitude and the $J/\psi N \to J/\psi N$ FSI term.
In addition to the dominant Pomeron-exchange contribution, meson exchanges in the
$t$-channel and direct $J/\psi$ radiation in the $s$- and $u$-channels are included
as part of the Born contribution.

We have extensively studied the role of light-meson and charmonium–meson exchanges.
Our analysis shows that the light mesons $\eta(548)$ and $\eta'(958)$ play a crucial
role in describing the available JLab data~\cite{GlueX:2019mkq,GlueX:2023pev,
Duran:2022xag}, while the contribution from charmonium mesons to the $\gamma p \to
J/\psi p$ reaction is negligible.
The direct $J/\psi$ radiation contribution is approximately two orders of magnitude
smaller than the meson-exchange contribution and becomes visible only in the
backward-angle region.
The hadronic form factors are introduced in a way that preserves gauge invariance.

For the $J/\psi N$ FSI term, the charmonium–nucleon potential is assumed to take a
Yukawa form [$-v_0\,\rm{exp}(-\alpha r)/r$], motivated by the gluon-exchange
interaction.
To examine the model dependence of the FSI effect, two different sets of parameters
are tested~\cite{Kawanai:2010ev,Brodsky:1989jd} .
The FSI effect is found to be essential for describing the threshold region when the
parameter set ($v_0$,\,$\alpha$) = (0.42,\,0.6 GeV)~\cite{Brodsky:1989jd} is used.
More precise, angle-dependent measurements in the near-threshold region
($E_\gamma \leqslant 8.9~\mathrm{GeV}$) are essential for verifying the FSI effect.

%=====================================================================================
\section*{Acknowledgments}
%=====================================================================================

The work was supported by the Basic Science Research Program through the National
Research Foundation of Korea (NRF) under Grant No. RS-2021-NR060129.

%=====================================================================================

%=====================================================================================

%-------------------------------------------------------------------------------------
\end{document}